\documentclass[a4paper]{JHEP3}

\usepackage{amsmath}
\usepackage{amssymb}
\usepackage{latexsym}
\usepackage{cite}
\usepackage[dvips]{graphicx}
\usepackage{bbold}
\usepackage{epsfig}
\usepackage{amsfonts}

\newcommand{\beqa}{\begin{eqnarray}}
\newcommand{\eeqa}{\end{eqnarray}}
\newcommand{\bea}{\begin{eqnarray}}
\newcommand{\eea}{\end{eqnarray}}

\newcommand{\be}{\begin{equation}}
\newcommand{\ee}{\end{equation}}
\newcommand{\half}{\frac{1}{2}}
\newcommand{\tr}{\,\textup{tr}}

\newcommand{\mcA}{{\mathcal A}}
\newcommand{\mcB}{{\mathcal B}}
\newcommand{\mcC}{{\mathcal C}}
\newcommand{\mcD}{{\mathcal D}}
\newcommand{\mcF}{{\mathcal F}}

\newcommand{\mcO}{{\mathcal O}}
\newcommand{\mcR}{{\mathcal R}}
\newcommand{\mcV}{{\mathcal V}}

\newcommand{\mcK}{{\mathcal K}}

\renewcommand{\Im}{\textup{Im}}

\newcommand{\bj}{\bar j}
\newcommand{\bi}{\bar i}
\newcommand{\bz}{\bar z}
\newcommand{\bw}{\bar w}

\newcommand{\bb}{\bar b}
\newcommand{\ewe}{\epsilon\wedge\bar{\epsilon}}

\numberwithin{equation}{section}

\title{Hermitian Yang-Mills instantons\\ on resolutions of Calabi-Yau cones}
\author{Filipe Paccetti Correia\\ Centro de F\' isica do Porto\\ Faculdade de Ci\^encias da Universidade do Porto\\ Rua do Campo Alegre, 687, 4169-007 Porto, Portugal\\ E-mail: \email{paccetti@fc.up.pt}}
\abstract{We study the construction of Hermitian Yang-Mills instantons over resolutions of Calabi-Yau cones of arbitrary dimension. In particular, in $d$ complex dimensions, we present an infinite family, parametrised by an integer $k$ and a continuous modulus, of SU($d$) instantons. A detailed study of their properties, including the computation of the instanton numbers is provided. We also explain how they can be used in the construction of heterotic non-K\"ahler compactifications.}

\preprint{}

\begin{document}

\section{Introduction}

Hermitian Yang-Mills (HYM) instantons play a central r\^ ole in supersymmetric heterotic compactifications, as some of their massless fluctuations should lead to the low energy standard model fields \cite{Candelas:1985en}. A great deal of progress has been done over the last two decades in the study of HYM instantons over (compact) Calabi-Yau 3-folds, mostly resorting to algebraic geometric techniques. This type of description, however, does not provide one with the instanton's hermitian connection, with the consequence that computing the (moduli dependent) values of the low energy couplings is inaccessible to analytical calculations. It is in fact expected that an exact knowledge of both the CY metrics and the hermitian connections is to remain out of reach. That having been said, one still can obtain partial information e.g. by looking at local aspects of heterotic compactifications, as we shall do here, or by focusing on certain points of moduli space and computing approximate CY metrics and HYM connections by means of numerical algorithms.

In the present paper we shall thus be interested on HYM instantons over a certain class of non-compact Ricci-flat K\"ahler manifolds, namely resolutions of Calabi-Yau cones with an arbitrary number of complex dimensions. Unlike the compact case, metrics on this class of non-compact Calabi-Yau's are explicitely known \cite{Calabi:1979,BB:1982,Page:1985bq,Candelas:1989js,Oota:2006pm,Lu:2006cw,Martelli:2007pv}, and it is the purpose of this work to find explicit HYM connections over them. The resolutions to be considered herein are a subset \cite{Calabi:1979,BB:1982,Page:1985bq} of the class of the canonical resolutions, for which the conical singularity gets replaced by a blown-up divisor. We shall leave HYM instantons over more general canonical resolutions as well as over small resolutions to future work.

The present paper is a natural continuation of research started with the construction of HYM instantons over CY cones in \cite{Correia:2009ri}. As the resolutions considered herein modify the CY geometry only near the tip of the cone, moreover without breaking the isometries of the background, it is natural to expect that the techniques and ans\"atze of \cite{Correia:2009ri} can be applied to the present problem without stark modification. This is indeed the case. By contrast, the computation of the instanton numbers
for the large class of solutions found in this paper turns out to be less trivial and hard to perform unless a trick presented in section \ref{sec:SU(3)solutions} is used.

As we said, the present work is motivated by heterotic compactifications. Indeed, the instantons presented herein can be used to construct heterotic non-K\"ahler \cite{Strominger:1986uh,Hull:1986kz} solutions with non-vanishing flux and varying dilaton in a way to be described in section \ref{subsec:nonkahler}. However, also from the viewpoint of gauge/gravity duality there is a natural interest in studying heterotic solutions beyond the standard embedding (see e.g. \cite{Carlevaro:2009jx}). The fact that in this case the geometry ceases being conical is of relevance, as in the dual gauge theory this corresponds to a departure of conformality. In a certain decoupling limit \cite{Aharony:1998ub}, these non-K\"ahler solutions become dual to gauge theories with massive flavor \cite{flavor_paper}.\\

The paper is organised as follows. We first introduce, in section \ref{sec:resolutions}, the geometry of Ricci-flat canonical resolutions of CY cones. Then, in section \ref{sec:instantons} we start our investigation of HYM instantons over these resolutions. We devote that section to a derivation of an equivalent problem formulated in terms of Higgs-YM flow equations. Several properties following from these equations, including expressions for instanton numbers, are obtained therein. An ansatz is then proposed in section \ref{sec:ansatz1} that reduces our problem to a simple one-dimensional non-linear differential equation. Implications for the computation of instanton numbers are presented. Also, we explain how to use solutions obeying this ansatz to construct heterotic non-K\"ahler compactifications with non-abelian instantons. Finally, in section\ref{sec:SU(3)solutions}, the solutions of the non-linear differential equation of the previous section are studied and classified, and for each solution the instanton numbers are computed. Appendix \ref{app A} contains a brief collection of useful results on CY cone geometry.

\section{Resolutions of CY cones}\label{sec:resolutions}

In this paper we shall consider a certain class \cite{Calabi:1979,BB:1982,Page:1985bq} of Ricci-flat resolutions of Calabi-Yau cones. We start with a review of the relevant properties of these resolutions and in the way introduce some useful notation.

Our ansatz for the K\"ahler 2-form $J=-ig_{a\bb}dz^a\wedge d\bz^{\bb}$, describing the canonical resolution of a CY cone of complex dimension $d=n+1$, is
\be
                   J=e^u J_{EK}-\frac{i}{2}e^v\epsilon\wedge\bar{\epsilon} \ ,
\ee
where $J_{EK}$ is the K\"ahler-form of an Einstein-K\"ahler n-fold, whose scalar curvature is determined to be $R_{EK}=4n(n+1)$ by the condition that the resolution is Ricci-flat (cf. the Appendix). We will sometimes use local holomorphic coordinates $z^a=\{z,w^i\}$, where the $w^i$ span the EK n-fold. The one-form
\be
                   \epsilon=\frac{dz}{z}+2\partial\mcK_{EK}
\ee 
is (locally) defined by the K\"ahler potential $\mcK$ on the EK n-fold, and $u$ and $v$ will be functions of the radial coordinate $t\equiv\ln2|z|^2+2\mcK_{EK}$. In terms of a n-bein on $EK_n$ (such that $J_{EK}=-ie_i\wedge\bar{e}_i$), one can introduce an holomorphic $(n+1,0)$-form, defined as
\be
                   \Omega=(ze^{\mcK_{EK}})^{n+1}\epsilon\wedge e_1\wedge e_2\wedge\dots\wedge e_n \ .
\ee

The condition that the resolution is Calabi-Yau translates into conditions on the forms $J$ and $\Omega$ that read
\be\label{eq:calibration_1}
                   dJ=0 \ ,
\ee
and
\be\label{eq:calibration_2}
                   i\Omega\wedge\bar{\Omega}\propto J^{n+1} \ .
\ee
From Eq.\eqref{eq:calibration_1}, it follows that
\be
                   \partial_t e^u=e^v \ .
\ee
It is then sensible to write the metric also in terms of a new radial coordinate $r^2=e^{u(t)}$. Introducing $f^2(r)=e^{v-u}$, we obtain
\be\label{eq:resolved_cone_new}
                   ds^2=\frac{dr^2}{f^2(r)}+r^2f^2(r)\eta^2+2r^2ds^2_{EK} \ ,
\ee
where $\eta=\Im~\epsilon$ is a real one-form satisfying $d\eta=-2J_{EK}$. Locally, we can write $\eta=d\psi-i(\partial-\bar{\partial})\mcK_{EK}$. From Eq.\eqref{eq:calibration_2} one finds that $u$ and $v$ must satisfy
\be
                   nu(t)+ v(t)-(n+1)t= \phi_0 \ ,
\ee
where $\phi_0$ is a constant, which in turn implies $f^2(r)$ to satisfy
\be\label{eq:dilaton_diffeq}
                  \partial_rf^2+2\frac{n+1}{r}(f^2-1)=0 \ .
\ee
This can then be integrated to give \cite{Calabi:1979,Page:1985bq}
\be\label{eq:f_lowest}
                 f^2(r)=1-\left(\frac{a^2}{r^2}\right)^{n+1} \ .
\ee
Notice that in the $a\to 0$ limit, we retrieve the (singular) CY cone. Let us also point out that Eq.\eqref{eq:dilaton_diffeq} follows from the condition that the K\"ahler resolution is Ricci-flat. In case we would drop the requirement of Ricci-flatness, $f^2(r)$ would be unconstrained.

Being Ricci-flat and K\"ahler, these resolutions can be used in (local) heterotic compactifications, provided one also takes care of the HYM instantons necessary to satisfy the anomaly cancellation conditions. We will see, in the following sections, how an infinite number of new families of such instantons can be constructed using the techniques of \cite{Correia:2009ri}. Once the backreaction of the HYM instantons is taken into account, in general the geometry ceases being K\"ahler and Ricci-flat. These effects kick in at next order in the $\alpha'$ expansion, and will be discussed in Section \ref{subsec:nonkahler} .

\section{HYM instantons on resolutions of CY cones}\label{sec:instantons}

We turn now our attention to the construction of (non-Abelian) Hermitian Yang-Mills instantons on the resolved CY cones of complex dimension $d_{\mathbb C}=n+1$, that is Yang-Mills instantons whose connections $A$ satisfy \cite{donaldson_1985,uhlenbeck:1986}
\be
                  \mcF_{2,0}(\mcA)=0 \ ,
\ee
\be
                  J^n\wedge\mcF_{1,1}(A)=0 \ .
\ee
Here, $\mcA$ is the $(1,0)$-part of the HYM connection $A=\mcA+\bar{\mcA}$, and we assumed the instanton to have a vanishing first Chern-class ($c_1(\mcF)=0$).\footnote{As usual in literature, we take the YM connection to be anti-hermitian, $A^{\dagger}=-A$, so that $\bar{\mcA}^{\dagger}=-\mcA$.} These two conditions are called holomorphy and Donaldson-Uhlenbeck-Yau (DUY) conditions, respectively.

To construct HYM instantons on the resolutions of CY cones discussed in the previous section we shall use the same ansatz that we introduced in our recent work \cite{Correia:2009ri} for instantons on sigular CY cones. That is
\be\label{eq:ansatz1}
                \mcA=\Phi(t,w,\bar{w})\epsilon+\mcB_i(t,w,\bar{w})dw^i \ ,
\ee
with $t$ being the radial coordinate defined above and the gauge choice $\Phi^{\dagger}=\Phi$. For this ansatz, the holomorphy and DUY conditions on the resolution of the CY cone read respectively
\be\label{eq:holom1}
             \mcF_{2,0}(\mcB)=0 \ ,
\ee
\be\label{eq:holom2}
             \partial_t\mcB=\mcD_{\mcB}\Phi \ ,
\ee
where $\mcD_{\mcB}=\partial+[\mcB,\cdot~]$, and
\be\label{eq:DUY1}
             \partial_u\Phi=\frac{1}{f^2}\partial_t\Phi=-n\Phi+\tfrac{1}{4}\mcK^{i\bar j}\mcF_{i\bar j}(B) \ .
\ee
Here and in the following, when convenient, we will interchange between the variables $t,~r$ and $u$. Since any of these variable is an increasing function of any of the others, we are save to choose any of them as the radial coordinate of the resolution of the CY cone.

By definition, instantons should have finite instanton numbers. In the present case this requires that $\partial_t\Phi=0=\partial_t\mcB$ at the boundaries located at $t=\pm\infty$. For $f^2=1$, i.e. on CY cones, we can set $u(t)=t$ and the above system of equations defines a Higgs-Yangs-Mills flow on the Einstein-K\"ahler $n$-fold \cite{Correia:2009ri}, which (for bounded instantons) interpolates between two fix points satisfying
\be
             \mcF_{2,0}(\mcB)=0 \ ,
\ee
\be
             \mcD_{\mcB}(\mcK^{i\bar j}\mcF_{i\bar j}(B))=0 \ ,
\ee
and
\be
             \Phi=\tfrac{1}{4n}\mcK^{i\bar j}\mcF_{i\bar j}(B) \ .
\ee
In \cite{Correia:2009ri} we called the $t=+\infty$-limit UV-limit, likewise calling $t=-\infty$ the IR-limit. Here, we will make use of this language for $u(t)\neq t$ too, keeping in mind that the structure of the HYM equations is modified in the IR.\footnote{A physical reason for this nomenclature, is that deformations of these instantons can be used also to construct the geometrical duals of certain gauge theories. As usual in gauge/gravity duality, the radial coordinate is dual to the energy scale of the gauge theory.}

\subsection{Entropy functional}

In \cite{Correia:2009ri} we found (for instantons on CY cones) that several properties of the Higgs-YM flow could be inferred by looking at a certain functional changing monotonically under the flow. A similar functional with this property can also be introduced for the present flow\footnote{We take $d\mu(EK_n)=\tfrac{1}{n!}(J_{EK_n})^n$ to be the volume form on the EK$_n$ base. Then, on the SE$_{2n+1}$ the volume form reads $d\mu(SE_{2n+1})=\eta\wedge d\mu(EK_n)$.}
\be\label{eq:def_M}
              N(t)=\frac{1}{Vol(EK_n)}(\partial_{u}+n)\int_{EK_n}\tr(\Phi^2) ~d\mu(EK_n) \ ,
\ee
which can easily be shown to satisfy 
\be\label{eq:for_M}
              \partial_t N(t)=\frac{1}{Vol(EK_n)}\int_{EK_n}\tr\left[\frac{2}{f^2}(\partial_t\Phi)^2+\mcK^{i\bj}D_i\Phi{\bar D}_{\bj}\Phi\right] ~d\mu(EK_n) \geq 0 \ ,
\ee
the equality holding only for $\partial_t\Phi=0=\partial_t\mcB$. Inspection of the flow equations shows that $N^{UV}\geq 0$. In case $\Phi^{IR}=0$, we find that $N^{IR}=0$, hence, as in the $a=0$ case discussed in \cite{Correia:2009ri}, unless the instanton is $t$-independent, $\Phi^{UV}\neq 0$. The opposite is also true. In case $\Phi^{UV}= 0$ we find that $N^{UV}=0$, implying that $\Phi^{IR}\neq 0$, otherwise we would have $N^{IR}=0$ and therefore no flow at all. These facts will be useful in Section \ref{sec:ansatz1}.

Within heterotic compactifications, the entropy functional turns out to have a physical meaning. To understand what this is, let us consider the heterotic 3-form flux $H_3$. This flux is induced both by the geometry and the gauge instanton. To $H_3$ we will associate a charge $q_{\infty}$ defined as
\be
               q_{\infty}=\lim_{t\to\infty} q(t) \ , \qquad q(t)=\frac{1}{\alpha'(n-1)!}\int_{SE(t)}H_3\wedge J_{EK}^{n-1} \ ,
\ee
where $q(t)$ should be interpreted as the charge inside the $(2n+1)$-dimensional Sasaki-Einstein manifold at radius $t$. Using the Bianchi identity
\be
               dH_3=\frac{\alpha'}{4}\left(\tr\mcR^2-\tr\mcF^2\right) \ ,
\ee
where $\mcR$ is the curvature 2-form on the resolved cone, we find that
\be\begin{split}\label{eq:charge}
               q(t) & =\frac{1}{8(n-1)!}\int_{SE(t)}\left(\tr\mcF^2-\tr\mcR^2\right)\wedge\eta\wedge J_{EK}^{n-2} \\
                    & =2\left(N_V(t)-N_S(t)\right)Vol(SE_{2n+1})+\frac{1}{8(n-1)!}\int_{SE(t)}\left(\tr\mcF_{B,V}^2-\tr\mcF_{B,S}^2\right)\wedge\eta\wedge J_{EK}^{n-2} \ .
\end{split}\ee
Here, $V$ denotes the gauge instanton (vector bundle) while $S$ stands for the spin-connection, which as will see below is also an instanton of the type described by our ansatz. Now, it is not difficult to see that $\partial_t\int\tr\mcF_B^{~2}\wedge J_{EK}^{n-2}=0$ follows from Eq.\eqref{eq:holom2}.\footnote{In fact, regarding $\mcB$ as being the connection of a holomorphic vector bundle $\mcV_B$ over the EK $n$-fold, Eq.\eqref{eq:holom2} states that $t$ is an instanton modulus for $\mcV_B$. In turn, this implies that varying $t$ cannot change a topological invariant such as the second Chern number of $\mcV_B$.} Hence, the 2nd term on the r.h.s. of \eqref{eq:charge} is a constant that can be evaluated at any value of the radial coordinate $t$. We shall see below that if at the IR the gauge instanton approaches the spin-connection, the flux is non-singular. Assuming this to be the case we find
\be
               q(t) =2\left(N_V(t)-N_S(t)\right)Vol(SE_{2n+1}) \ .
\ee
That is, the functional $N(t)$ determines the charge inside the $(2n+1)$-dimensional Sasaki-Einstein manifold at radius $t$.

Finally, for later use let us note that the entropy functional can be expressed in terms of the second Chern class of the instantons as follows
\be\label{eq:L_A(t)}
             L_{\mcA}(t)\equiv\int_{EK_n}\tr\mcF^2_A\wedge J_{EK}^{n-2} =16(n-1)!N(t)Vol(EK_n)+\int_{EK_n}\tr\mcF_B^{~2}\wedge J_{EK}^{n-2} \ .
\ee
 This means that also
\be
            \partial_t L_{\mcA}(t) \geq 0 \ ,
\ee
under the flow.

\subsection{Instanton numbers}\label{sec:inst_numbersI}

In $d$-complex dimensions, any instanton is characterised by a topological charge, the so-called instanton number $N_d$, that is constructed by integrating the $d$-th Chern character over the CY d-fold
\be
                   N_d=\frac{1}{d!}\int\tr\left(\frac{i\mcF}{2\pi}\right)^d \ .
\ee 
We shall now derive expressions for the instanton numbers in $d=2,3$ complex dimensions. Notice that due to the topological nature of $N_d$ no use of the DUY equation is required, only the holomorphy condition will play a role in the following computations.

Consider first instantons over CY cone 2-folds ($n=1$). It is straightforward to find that the instanton number,
\be
                   N_2=\frac{1}{\pi^2}\int^{UV}_{IR}(\partial_u+1)\tr\Phi^2\eta\wedge J_{EK} \ ,
\ee
is closely related with the entropy functional introduced above.

For instantons over CY cone 3-folds ($n=2$), we find that
\be\begin{split}
                   \tr\mcF_A^3 = & \partial_t\left[32\tr\Phi^3 J_{EK}^2 +24i\tr(\Phi^2\mcF_B)\wedge J_{EK}-6\tr(\Phi\mcF_B^2)\right]\wedge\ewe\\
                               & +\partial\bar{\partial}G_{1,1}\wedge\ewe \ ,
\end{split}\ee
where 
\be
                   G_{1,1}=-16i\tr\Phi^3 J_{EK}-12\tr(\Phi^2\mcF_B) \ ,
\ee
is a $(1,1)$-form on the EK 2-fold. Assuming the latter to be compact, it is then possible to write down the instanton number $N_3$ in terms of quantities defined on the EK base. It is useful to introduce the following set of gauge invariant forms
\be
                   C_{2k}^{(3)}(\Phi,B)=\frac{1}{k!}\tr\left[\Phi^{3-k}\left(\frac{i\mcF_B}{2\pi}\right)^k\right] \ .
\ee
We then find that
\be
                   N_3=-\frac{2}{3\pi^3}\int\tr\Phi^3\Big{|}_{IR}^{UV} J_{EK}^2\wedge\eta-\frac{1}{\pi^2}\int C_2^{(3)}\Big{|}_{IR}^{UV}\wedge J_{EK}\wedge\eta-\frac{1}{\pi}\int C_4^{(3)}\Big{|}_{IR}^{UV}\wedge\eta \ .
\ee
We expect, but do not prove, that for any number of dimensions the instanton numbers can be expressed in terms of invariants constructed by tracing over products of powers of $\Phi$ and $\mcF_B$. 

In three complex dimensions $(n=2)$, the instanton is in addition characterised by the integral of the second Chern character over the Einstein-K\"ahler base at $t=-\infty$, for the latter defines a compact four-cycle. As follows from \eqref{eq:L_A(t)}, this reads
\be
                   \int_{EK_2}ch_2(\mcF_A)=-\frac{2}{\pi^2}N(-\infty)Vol(EK_2)+\int_{EK_2}ch_2(\mcF_B) \ .
\ee

\section{SU(n+1) instantons}\label{sec:ansatz1}

We specialise now to SU(n+1) instantons, $n$ being the number of complex dimensions of the Einstein-K\"ahler base. We shall assume firstly that the adjoint Higgs $\Phi$ is constant over the EK$_n$ base, secondly that it points always in the same direction in the gauge symmetry space, that is 
\be
             \Phi(t)=\varphi(t)\Sigma \ , \quad \Sigma=\left(\begin{array}{cc}
              1 &\quad 0 \\
              0 & -{\mathbb 1}_n
               \end{array} \right) \ . 
\ee
Writing the connection $\mcB$ on the EK n-fold as
\be
              \mcB=\left(\begin{array}{cc}
              \alpha &\quad E_{-} \\
              E_+ & \Gamma_{n\times n}
              \end{array} \right) \ ,
\ee
Eq.\eqref{eq:holom2} then implies that $\partial_t \alpha=0=\partial_t\Gamma$, while
\be
            \partial_t E_{\pm}=\pm\frac{n+1}{n}\varphi(t) E_{\pm} \ ,
\ee
and thus
\be
            E_{\pm}(t,w,\bw)=e^{\pm\frac{n+1}{n}\int^t \varphi dt}E_{\pm}^0(w,\bw) \ .
\ee
Now, suppose that both $E_+^0$ and $E_-^0$ are non-vanishing. Then, the condition that $\partial_t\mcB=0$ at the IR and the UV would imply that $\varphi=0$ both at the IR and the UV. However, as we pointed out above, \eqref{eq:for_M} would then imply the instanton to be a $t$-independent HYM instanton on the EK$_n$ base space and thus of no interest to us. We shall therefore take $E_-^0=0$, denoting $E_+=E$ from now on. Setting instead $E_+^0=0$ would simply amount to a different gauge choice.

Getting back to the condition that $\mcB$ be holomorphic ($\mcF_{2,0}(\mcB)=0$), we learn that
\be
           \partial\alpha=0 \ , \qquad \partial\Gamma+\Gamma\wedge\Gamma=0 \ ,
\ee
and
\be
           \partial E^0+E^0\wedge\alpha+\Gamma\wedge E^0=0 \ .
\ee
This can be solved \cite{Correia:2009ri} by setting $\alpha=\partial\mcK$, $\Gamma=\mcC+{\mathbb 1}_n\partial\mcK$ and $E^0_i=-\sqrt{2}e_i$, where $\mcC$ is the spin-connection on the EK$_n$ base (cf. Eq.\eqref{eq:spin_connec_cone}). With this choice, the gauge curvature of $B$ reads
\be\label{EQ:F_B}
                   \mcF(B)=\mcR_C+2\left(\exp\left(2\frac{n+1}{n}\int\varphi~dt\right)-1\right)\left(\begin{array}{cc}
              e_i\wedge\bar{e}_i & ~0 \\
              ~0 & -e_a\wedge\bar{e}_b
              \end{array} \right) \ ,
\ee
where $\mcR_C$ is the CY cone curvature 2-form, and the DUY condition becomes
\be\label{eq:DUY_varphi}
           \partial_u\varphi+n\varphi+\frac{n}{2}\left(1-\exp\left(2\frac{n+1}{n}\int^t_{t_0}\varphi dt\right)\right)=0 \ .
\ee
This equation will be the object of a detailed study later in Section \ref{sec:SU(3)solutions}, where we will find a family of solutions parametrised by an integer $k$ and a continuous modulus, distinct $k$'s denoting topologically distinct instantons. Before studying the solutions of this equation, let us compute the instanton numbers, and discuss the implications of our ansatz for heterotic compactifications on resolved cones.

\subsection{Instanton numbers}

We can evaluate the instanton numbers by directly using the results of Section \ref{sec:inst_numbersI} but it might be enlightening to have explicit expressions for the Chern characters too. The instanton's curvature 2-form $\mcF(A)$ reads
\be
                   \mcF(A)=\mcF(B)-2\partial_t\Phi\epsilon\wedge{\bar\epsilon}-4i\Phi J_{EK}-\partial_t\mcB\wedge{\bar\epsilon}+\epsilon\wedge\partial_t{\bar\mcB} \ ,
\ee
with $\mcF(B)$ as given in \eqref{EQ:F_B}. For the $n=1$ case we then easily find that
\be\label{eq:N_2}
                \tr\mcF^2_A=16\partial_t\left(\varphi+\varphi^2-\varphi e^{4\int\varphi}\right)dt\wedge\eta\wedge J_{EK} \ ,
\ee
while for the $n=2$ case we have
\be
                \tr\mcF^3_A=-3id\varphi\wedge\eta\wedge\tr\mcR^2_{EK}-12idL\wedge\eta\wedge J^2_{EK} \ ,
\ee
with
\be\label{eq:def_L}
                L(t)=6\varphi+3\varphi^2+2\varphi^3-3\varphi^2 e^{3\int^t\varphi}-6\varphi e^{3\int^t\varphi}+3\varphi e^{6\int^t\varphi} \ .
\ee
Clearly, up to topological invariants of the background geometry, the instanton numbers are completely determined by the behaviour of $\varphi(t)$ at the IR and the UV boundaries. Due to the powers of $e^{\int^t\varphi}$ appearing in Eqs.\eqref{eq:N_2} and \eqref{eq:def_L}, finiteness of $N_2$ and $N_3$ requires that
\be
                -\infty<\varphi_{UV}\leq 0\leq\varphi_{IR}<+\infty \ .
\ee
Furthermore, inspection of the DUY equation shows that both at the IR and UV, $e^{\int^t\varphi}\to const$. We thus find
\be\label{eq:N_2_new}
               N_2=-\frac{2}{\pi^2}(\varphi+\varphi^2)^{UV}_{IR}\int_{SE_3} \eta \wedge J_{EK} \ ,
\ee
and
\be\label{eq:N_3_new}
               N_3=\frac{\varphi_{UV}-\varphi_{IR}}{4\pi}\int_{SE_5}\eta\wedge\tr\left(\frac{i\mcR_{EK}}{2\pi}\right)^2-\frac{1}{4\pi^3}\left(6\varphi+3\varphi^2+2\varphi^3\right)^{UV}_{IR}\int_{SE_5}\eta\wedge J_{EK}^2 \ .
\ee
In Section \ref{sec:SU(3)solutions} we will study in detail the solutions of the DUY equation. For each solution, we will be able to find explicit values for $\varphi_{IR}$ and $\varphi_{UV}$ thus determining the instanton numbers (up to the topological invariants of the underlying Sasaki-Einstein geometry).

\subsection{Heterotic non-K\"ahler compactifications}\label{subsec:nonkahler}

We describe now how to solve the heterotic BPS equations \cite{Strominger:1986uh,Hull:1986kz} to first order in $\alpha'$ in three complex dimensions (known as the Strominger system of equations), taking the configuration to be an SU(3) instanton of the type described above over a resolved CY$_3$ cone. One should proceed as follows. First we solve the HYM equations on the CY cone background. Then, on the instanton background we compute the non-K\"ahler corrections to the geometry, the dilaton and also the induced flux, all of them at order $\alpha'$. Finally, the $\alpha'$-correction to the instanton is accounted for by solving the DUY equation \eqref{eq:DUY_varphi} on the new background but we shall not consider this here.

The non-K\"ahler deformation of the resolved cone is determined by solving
\be
              i2\partial\bar{\partial}J=\frac{\alpha'}{4}\left(\tr\mcF\wedge\mcF-\tr\mcR\wedge\mcR\right) \ .
\ee
Now, our ansatz implies that
\be\label{eq:2ndchernclass}
             \tr\mcF^2=\tr\mcR_C^2+2\partial\bar{\partial}\left[(\partial_u+2)\tr\Phi^2\epsilon\wedge\bar{\epsilon}\right] \ .
\ee
As we shall see below, also the spin-connection of the resolved CY cone can be written as an SU(3) HYM instanton of the type described in this section. As such, the first Pontryagin class of the tangent bundle is also of the form \eqref{eq:2ndchernclass}. Then, we easily find that, to first order in $\alpha'$, the geometry is given by
\be\label{eq:geo_1st_order}
            J=J_{CY}+i\frac{\alpha'}{4} (N_S-N_V)\,\epsilon\wedge\bar{\epsilon} \ ,
\ee
where $J_{CY}$ is the (lowest order) Calabi-Yau background, and the subscripts $S$ and $V$ are used to denote the contributions of the spin-connection and vector bundle, respectively, to the non-K\"ahler deformation of the geometry. Notice that, for the present ansatz, the entropy functionals are given as $N_{V,S}=(\partial_u+2)\tr\Phi_{V,S}^2$. It is now straightforward to obtain also the three-form flux $H$:
\be 
             H=i(\partial-\bar{\partial})J=-\alpha'(N_S-N_V)\,\eta\wedge J_{EK} \ .
\ee
Finally, we find that the dilaton should obey
\be\label{eq:phi_1_order}
             \partial_r e^{\phi}=-\frac{\alpha'}{f^2(r)r^3}(N_S(r)-N_V(r)) \ .
\ee
Together with Eq.\eqref{eq:dilaton_diffeq}, this equation determines the geometry and the dilaton to first order in $\alpha'$. That means that, on the R.H.S. of \eqref{eq:phi_1_order}, one should set $f=1-(a/r)^6$, and the instanton should be computed on this background too. 

Now, $r^3f^2(r)$ vanishes at the apex of the resolved cone (i.e. at $r=a$), including for $a=0$. It follows that $(N_S-N_V)$ should also vanish there, otherwise $\partial_r\phi$ would be singular at the IR and we would not be able to trust our solution in that region. This condition is equivalent to the requirement that for small $r$
\be
                 \int_{EK_2,~r=a}(\tr\mcF^2-\tr\mcR^2)=0 \ ,
\ee
or that the flux three-form $H$ should be well-defined at the apex of the resolved cone. 

Looking now at the behaviour at the UV let us point out that, as we shall see below, $\Phi_S\to 0$ for $r\to+\infty$. This is expected as for $r \gg a$ the geometry becomes that of a CY$_3$ cone (i.e. $f^2=1$) and we know \cite{Correia:2009ri} that on a cone $\Phi_S=0$ everywhere. Then, at the UV we find
\be
                 \partial_r e^{\phi}\simeq-\frac{\alpha'}{r^3}N_V(r) \ .
\ee
Since we also know \cite{Correia:2009ri} that on the cone $N_V(r)\geq 0$, we can conclude that for $r\gg a$, the heterotic string coupling $e^{\phi}$ is a decreasing function of the radius $r$. In fact, for $r\to +\infty$ we find
\be
                 e^{\phi}\to e^{\phi_{\infty}}+\frac{3}{8}\frac{\alpha'}{r^2} \ .
\ee
This simple behaviour is modified at values of $r$ close to the instanton's size and/or close to the size of the resolution, $a$. In particular, the strength of the string coupling at the IR is set by the value of the instanton and/or K\"ahler moduli. Then, weak coupling is guaranteed by taking large values of these moduli and the dilaton modulus $e^{\phi_{\infty}}$ sufficiently small.

NSNS flux charges are determined by the asymptotic behaviour of $H$ as \cite{Carlevaro:2009jx}
\be
                 Q=\lim_{r\to\infty}\frac{2}{\pi^2\alpha'}\int_{S^3} H \ ,
\ee
where the integral should be taken over asymptotic 3-spheres. Recall that the NSNS flux charges can be interpreted as the charges of fivebranes wrapping two-cycles at the IR. One can define asymptotic $S^3$'s by choosing some $\mathbb{CP}^1$ on the Einstein-K\"ahler base of the five-dimensional Sasaki-Einstein and pulling back this $\mathbb{CP}^1$ to an U(1) bundle over it the same way the EK two-fold is pulled-back to the SE space. Then,
\be
                 Q_i=\frac{4}{\pi^2}\tr\Phi^2_V(+\infty)\int_{S^3_i} \eta\wedge J_{EK} \ ,
\ee
where $i$ denotes the $\mathbb{CP}^1$ we are considering. From this we can learn two things. The first is that a non-vanishing $\tr\Phi^2_V(+\infty)$ is a necessary condition for non-vanishing flux charge. The second is that, our ansatz can only describe \emph{democratic} fivebrane configurations, that is fivebranes wrapping all $\mathbb{CP}^1$'s on the EK$_2$ base. As we will see in Section \ref{sec:SU(3)solutions}, SU(3) instanton solutions with non-vanishing $\Phi$ in the UV, necessarily have $\Phi(+\infty)=-\half\Sigma$. In that case
\be
                Q_i\to\frac{3}{2\pi^2}\int_{S^3_i} \eta\wedge J_{EK} \ .
\ee
Taking our three-fold to be the resolution of the cone over $T^{1,1}/\mathbb{Z}_2$, one would find that $Q_i=2/3$ for any of those SU(3) instantons. Instanton configurations with non-fractional charges will be presented in a companion paper \cite{flavor_paper}. There, we will construct 2 types of SU(2) instantons on the resolution of the cone over $T^{1,1}/\mathbb{Z}_2$, which interchange under an interchange of the two $\mathbb{CP}^1$'s on the base of that geometry. There we show that for one type of instanton $Q_1=3/2$, $Q_2=-1/2$, for the other $Q_1=-1/2$, $Q_2=3/2$. Then, a symmetric combination of $n$ SU(2) instantons of each type, such that the isometries of the geometry are respected, has $Q_1=Q_2=n$, and should be interpreted as a $n$ pairs of fivebranes wrapping the two $\mathbb{CP}^1$'s.

\section{Non-abelian instanton solutions}\label{sec:SU(3)solutions}

Here, we shall study and classify the solutions of the differential equation
\be\label{eq:DUY_varphi_n}
           f^{-2}\partial_t\varphi+n\varphi+\frac{n}{2}\left(1-\exp\left(2\frac{n+1}{n }\int^t_{t_0}\varphi dt\right)\right)=0 \ .
\ee
We shall keep the discussion as general as possible, before specialising to the $n=1,2$ cases and specifying the geometry dependent function $f(t)$. Indeed, for topological considerations, the precise form of $f(t)$ does not matter, only its behaviour at the IR and the UV boundaries. This means in particular, that for the purpose of a topological classification using a tree-level or an $\alpha'$-corrected $f(t)$ leads to the same result.

To get some insight on the solutions of Eq.\eqref{eq:DUY_varphi_n}, it is useful to recast it as
\be\label{eq:mec_analog}
          \partial_{\tau}^2X+(f+\partial_{\tau}\ln f)\partial_{\tau}X=-\frac{d}{dX}\left(X-Me^{X/M}\right) \ ,
\ee
where $X=2n\int^t\varphi$, $M=n^2/(n+1)$, and $\partial_t \equiv nf\partial_{\tau}$. Equation \eqref{eq:mec_analog} describes the damped motion of a particle with position $X$ in the potential $V(X)=X-Me^{X/M}$. Depending on the cone being resolved or singular, the IR geometry is described by
\be\label{eq:f^2}
             f^2\simeq\left\{\begin{array}{cc}
              c^2~\tau^2+\mcO(\tau^4) & , \textup{ resolved} \\
               1+a~e^{b\tau}& , \textup{ singular} 
              \end{array} \right. 
\ee 
with $\tau\geq 0$ in the resolved cone and $\tau\in\mathbb{R}$ in the singular one. In this paper we are interested in the former case, for which we see that the damping force diverges at $\tau=0$ unless $X'(0)=0$. Then, classifying the instantons with regard to their small $\tau$ behaviour, we easily find \emph{two} branches of relevant solutions:
\begin{itemize}
\item[(A)] Starting at $X(0)=-\infty$ with $X'(0)=+\infty$. This family of solutions includes the spin-connection of the resolved CY cone.

\item[(B)] Starting at finite $X(0)\leq 0$ with $X'(0)=0$. Taking the singular cone limit, this family of solutions becomes the one constructed in \cite{Correia:2009ri}.
\end{itemize}

%

\paragraph{A-branch.} Let us start by discussing the A-branch of solutions. Inspection of the Eq.\eqref{eq:mec_analog} shows that at small $\tau$ the solutions satisfy
\be\label{eq:A-branch_small}
          X\simeq X_0+k\frac{2n^2}{n+1}\ln\tau+{\mcO}(\tau^2) \ .
\ee
Imposing that $\varphi(\tau)$ be $C^{\infty}$ at $\tau=0$ we find $k\in{\mathbb N}^{+}$ to be a quantised parameter. By contrast, $X_0$ is a priori a continuous arbitrary real number. As we have seen above, the instanton number is (in part) determined by the value of $\varphi$ at the IR ($\tau=0$), 
\be
         \varphi_{IR}=c\frac{kn^2}{n+1} \ .
\ee
Through the constant $c$ (cf. \eqref{eq:f^2}), this depends also on the way $f^2$ increases in the IR. In case the geometry is CY, we found that $f^2(r)=1-\left(\frac{a^2}{r^2}\right)^{n+1}$, so that
\be
         f=\tanh\left(\frac{n+1}{2n}\tau\right) \ .
\ee
This means that on CY resolved cones, we have
\be
         \varphi_{IR}=\frac{nk}{2} \ .
\ee
Notice that the instanton number depends on $k$ but not on $X_0$, that meaning that $X_0$ is an instanton modulus. As we have already pointed out, the spin-connection of the CY resolved cone also belongs to this family of HYM instantons. It is described by
\be
          X_{S}=\frac{2n^2}{n+1}\ln\tanh\left(\frac{n+1}{2n}\tau\right)=M\ln f^2(\tau) \ ,
\ee
which has $k=1$ and $X_0=\frac{2n^2}{n+1}\ln\frac{n+1}{2n}$. Clearly, $X_S(\tau)$ describes the motion of a particle starting at position $X_S=-\infty$ with infinite positive initial velocity $X_S'=+\infty$. The particle goes \emph{up} to the maximum of the potential $V(X)$ at $X_S=0$, which it reaches at $\tau=+\infty$.

On the resolved CY cone, for given $k\in{\mathbb N}^{+}$, we expect a one-parameter family of solutions with the small-$\tau$ behaviour as in \eqref{eq:A-branch_small} to exist. To justify this expectation, we shall give an intuitive argument that goes as follows. Consider for $n=1$ (for general $n$ the argument is the same, only the mathematical expressions differ) the following \emph{approximate} solution: 
\be\label{eq:approx_1}
             X\approx X_0+k\ln\sinh\tau-(k+1)\ln\left(\tfrac{1}{2}+\tfrac{1}{2}\cosh\tau\right) \ .
\ee
This solution describes the motion of a particle that, starting at $X=-\infty$, never reaches the maximum of the potential. In fact, it attains a maximum value at finite time
\be
             \tau_{max}= \text{arc\,cosh}(k+1) ,
\ee 
where it turns and slides back to $X=-\infty$. This approximate solution can only be trusted if $X_{max}\equiv X(t_{max})\ll -1$, and this can be arranged by taking $X_0$ suficciently negative. The upshot of this discussion is that, using continuity arguments, we expect that, for given $k$, there should be a one-parameter family of solutions, characterised by the value of $X_{max}$ (or $X_0$), if $X_{max}\leq 0$. Clearly, of these solutions only one will have $X_{max}=X(\infty)=0$ as is the case with the spin-connection. In other words:
\begin{itemize}
\item [(I)] For $X_0$ ranging between $-\infty$ and a certain $k$-dependent maximal value $X_k$, our solutions will have $X(+\infty)=-\infty$;
\item [(II)] When $X_0=X_k$, our solution interpolates between $X(-\infty)=-\infty$ and $X(\infty)=0$. For $k=1$, this is the spin-connection;
 \item [(III)] For $X_0>X_k$, the solutions have enough energy to overcome the top of the potential and slide down to $X=\infty$. These solutions lead to unbounded instantons, and are therefore of no interest to us.
\end{itemize}
As an aside, we note that for large $k$ one can use the above approximate solution to obtain a fairly good estimate of $X_k$ by setting $X(\tau_{max})=0$:
\be
                  X_k\approx-\frac{k}{2}\ln(k^2 + 2k) + (k + 1)\ln(1 + k/2) \ .
\ee
Let us also remark that in the limit $X_0\to-\infty$, our SU(2) instantons become S[U(1)$^2$] HYM instantons.

\EPSFIGURE[p]{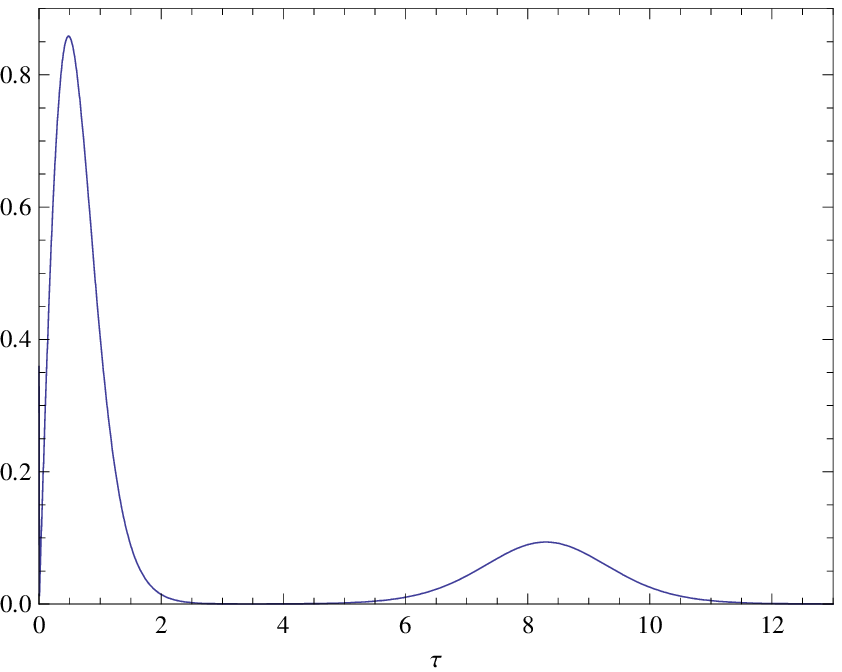,width=204pt}{Plot of the \emph{effective} instanton number density, obtained by multiplying the instanton number density by the area of the surface of constant $\tau$, as a function of $\tau$. The instanton depicted here has $n=1$, $k=1$, and $X_0=-10^{-3}$. One can observe how a part of the instanton left from the core of the instanton at apex of the resolved cone. Eventually, as $X_0\to 0$ this fraction of the instanton approaches infinity and leaves the geometry.\label{fig:1}}

What we have just stated for $n=1$ can be easily generalised to $n\neq 1$, in particular to $n=2$. The solutions of the A-branch fall into distinct topological classes characterised by an integer $k\geq 1 $. There is a second - continuous - parameter $-\infty<X_0<X_k$ that is a modulus connecting different instantons within the same topological class. At a certain critical value $X_0=X_k$, the instanton undergoes a topological transition. Since the transition happens at the UV, it cannot be a small instanton transition. In fact, what is happening is that when $X_0$ is close to the critical value $X_k$, the instanton breaks into two parts, one sitting at the apex, the other moving towards the UV as we approach the critical point. In the opposite limit, $X_0\to-\infty$, solutions in the A-branch become S[U(1)$\times$ U(n)] HYM instantons.

\paragraph{B-branch.} The solutions belonging to the B-branch are simple to characterise. They behave at small $\tau$ as
\be
          X\simeq -|\tilde{X}_0|-\frac{1}{4}(1-e^{-|\tilde{X}_0|})\tau^2 \ ,
\ee
and describe the motion of particles that start at a finite distance from the maximum of the potential and slide down towards $X=-\infty$. As above, for large $|\tilde{X}_0|$ one can write down an approximate solution valid for all values of $\tau$. For $n=1$ this reads
\be
             X\simeq -|\tilde{X}_0|-\ln\left(1+\cosh\tau\right) \ ,
\ee
and is clearly obtained by setting $k=0$ in Eq.\eqref{eq:approx_1}.

In two complex dimensions (i.e. $n=1$), the resolved CY cone is the Eguchi-Hanson \cite{eguchi} space obtained by blowing up a $\mathbb{CP}^1$ at the singularity of the $\mathbb{C}^2/\mathbb{Z}_2$ orbifold. The instanton number, computed according to \eqref{eq:N_2_new} reads
\be
                  N_2  =\half(k+1)^2-\half\delta_{\textup{II}} \ ,
\ee
where $\delta_{\textup{II}}=1$ if the instanton belongs to the A$_{\textup{II}}$-branch and $\delta_{\textup{II}}=0$ otherwise. The B-branch instantons have $k=\delta_{\textup{II}}=0$. We should emphasize that, from this family of instantons, sofar only the latter and the $k=\delta_{\textup{II}}=1$ instanton (i.e. the spin-connection) were explicitely known (see \cite{Bianchi:1996zj}).

Moving to three complex dimensions (i.e. $n=2$), the computation of the instanton number, constructed from the third Chern character, is also straightforward to perform using Eq.\eqref{eq:N_3_new}. It is however useful to consider another approach to this computation, that turns out to be better suited to being extended to an arbitrary number $n$ of dimensions. What we shall do is to explore the fact that in the limit that the modulus $X_0$ is close to the critical value $X_k$, the instantons of the A$_{\textup{I}}$-branch can be regarded as superpositions of an instanton on the A$_{\textup{II}}$-branch with the same value of $k$, localised at $\tau<\tau_k$, and an instanton belonging to the B-branch and located at $\tau\gg\tau_k$, as illustrated in Fig.\ref{fig:1}. This tells us that knowing the instanton numbers in the A$_{\textup{I}}$-branch and B-branch we can compute the instanton numbers in the A$_{\textup{II}}$-branch using simple subtraction
\be\label{eq:superimpose}
                N_3(\textup{A}_{\textup{II}})=N_3(\textup{A}_{\textup{I}})-N_3(\textup{B}) \ .
\ee
Crucially, the instanton numbers on the A$_{\textup{I}}$-branch and B-branch can be easily computed by using approximate solutions valid for $|X_0|\gg 1$ and $|\tilde{X}_0|\gg 1$, respectively. Indeed, for $n=2$ we find that in the A$_{\textup{I}}$-branch the instanton number is
\be\label{eq:N3_AIbranch}
                N_3(\textup{A}_{\textup{I}})=-\frac{2k+1}{8\pi}\int\eta\wedge\tr\left(\frac{i\mcR_{KE}}{2\pi}\right)^2+\frac{L_k}{4\pi^3}\int\eta\wedge J_{KE}^2 \ ,
\ee
where $L_k=5/2+6k+3k^2+2k^3$ and $\mcR_{KE}$ is the curvature two-form on the the Einstein-K\"ahler 2-fold. Moreover, the instanton number on the B-branch is retrieved by setting $k=0$ in \eqref{eq:N3_AIbranch}. This implies that on the A$_{\textup{II}}$-branch one should have
\be\label{eq:N3_AIIbranch}
                N_3(\textup{A}_{\textup{II}})=-\frac{k}{4\pi}\int\eta\wedge\tr\left(\frac{i\mcR_{KE}}{2\pi}\right)^2+\frac{L_k-5/2}{4\pi^3}\int\eta\wedge J_{KE}^2 \ .
\ee
The validity of this chain of arguments is supported by the agreement with the exact result obtained using \eqref{eq:N_3_new}.

Finally, let us comment on the UV asymptotics of our solutions. Since at the UV the geometry becomes that of a CY cone, our solutions are asymptotic to those discussed in \cite{Correia:2009ri}. This means in particular that $\varphi(+\infty)= -\half\textup{ or } 0$. Hence, solutions on the A$_{\textup{I}}$ and B-branches have $N(+\infty)=(n+1)/4$, while on the A$_{\textup{II}}$ branch have $N(+\infty)=0$.

\subsection{Instanton numbers for general $n$}

To compute the instanton numbers at arbitrary $n$ we employ the same line of reasoning as above. The idea is to compute $N_d$ for instantons on the A$_{\textup{I}}$ and B-branches in the limits $X_0\to -\infty$ and $\tilde{X}_0\to -\infty$, and then use \eqref{eq:superimpose} to obtain the instanton numbers also on the A$_{\textup{II}}$-branch. The point is that in such a limit, the gauge curvature simplifies to
\be
               \mcF_A=-2\partial_t\Phi\epsilon\wedge\bar{\epsilon}+\mcR_C-4\Phi e_i\wedge\bar{e}_i-2\left(\begin{array}{cc}
              e_i\wedge\bar{e}_i & ~0 \\
              ~0 & -e_a\wedge\bar{e}_b
              \end{array} \right) \ .
\ee
Then,
\be\begin{split}
              \mcF_A^{n+1} = & -2^n(-i)^{n+1}\partial_t(1+2\varphi)^{n+1}dt\wedge\eta\wedge J_{EK}^n\\
                           & -2i\frac{n+1}{n}\partial_t\varphi dt\wedge\eta\wedge\tr\left(\mcR_{EK}+i\mathbb{1}\left(\tfrac{4}{n}\varphi-2\right)J_{EK}\right)^n \ ,
\end{split}\ee
which clearly can easily be writen as a total derivative for any value of $n$. We just need to expand the second line of the equation in powers of $J_{EK}$ and $\mcR_{EK}$ to find that
\be
              N_{n+1}=\sum_{p=0}^n A_p\int_{SE_{2n+1}}\eta\wedge J_{EK}^{p}\wedge\tr\left(\frac{i\mcR_{EK}}{2\pi}\right)^{n-p} \ ,
\ee
with
\be
              A_p=-\frac{(1-2\varphi/n)^{p+1}}{2^{p-n+1}\pi^{p+1}(p+1)!(n-p)!}\bigg|_{IR}^{UV} \ ,
\ee
for $p\neq n$, and
\be
              A_n=-\frac{(1+2\varphi)^{n+1}+(1-2\varphi/n)^{n+1}}{2\pi^{n+1}(n+1)!}\bigg|_{IR}^{UV} \ .
\ee
The instanton number $N_{n+1}(k)$ valid for any instanton on the A$_{\textup{I}}$ and B-branches then follows by setting $\varphi_{UV}=-\half$ and $\varphi_{IR}=nk/2$, while on the A$_{\textup{II}}$ branch we just have to use that $N_{n+1}(\textup{A}_{\textup{II}})=N_{n+1}(k)-N_{n+1}(0)$.

\section{Outlook}

In this paper we studied and presented a large family of solutions of the HYM equations over certain canonical resolutions of CY cones of arbitrary complex dimension. This family of HYM instantons is parametrised by a integer $k$ and a single modulus parameter. We also developed the techniques necessary to compute their instantons numbers, which we showed to depend only on $k$ and the topology of the underlying geometry. The instantons presented herein can be used not only to construct local (non-K\"ahler) heterotic compactifications as we described, but also to construct gravity duals of certain supersymmetric gauge theories with massive flavor in a way to be explained in \cite{flavor_paper}.

Several avenues are open for further research. We would like to mention the task of computing the moduli spaces of the instantons presented in this paper and finding their K\"ahler structure. In particular, it would be nice to see how they behave as we send the blow-up modulus of the resolution to zero and the CY becomes a cone. This could be of relevance for the question of moduli stabilisation in heterotic compactifications, as emphasised in \cite{Correia:2007sv}. In \cite{Correia:2009ri} we speculated that in the vanishing blow-up modulus limit the instanton moduli space should be itself a K\"ahler cone. This is still an open issue.

\acknowledgments{The author is grateful to S. Groot Nibbelink and C. A. R. Herdeiro for useful discussions. This work is supported by \emph{Funda\c c\~ao para a Ci\^encia e a Tecnologia} through the grant SFRH/BPD/20667/2004, and the projects PTDC/FIS/099293/2008 and CERN/FP/ 109306/2009.}

\appendix

\section{Calabi-Yau cones}\label{app A}

In this appendix, we setup some notation and collect several useful results on Calabi-Yau cones. We start by recalling that a K\"ahler metric can always be locally written as
\be
             ds^2=2\mcK_{i\bj}~dz^i\otimes d\bz^{\bj} \ ,
\ee
where $\mcK_{i\bj}=\partial_i\partial_{\bj}\mcK(z,\bz)$ and $\mcK(z,\bz)$ is the K\"ahler potential. Clearly, this implies that the K\"ahler 2-form,
\be
             J=-i\mcK_{i\bj}~dz^i\wedge d\bz^{\bj} \ ,
\ee
is closed. Let us introduce an orthonormal frame
\be
             e_i=g^{\dagger}_{ij}dz^j \ , 
\ee
defined in terms of a squared matrix $g$ of rank $d_{\mathbb C}$ satisfying $(gg^{\dagger})_{\bj i}=\mcK_{i\bj}$, such that now
\be
             ds^2=2 e_i\otimes {\bar e}_{\bi} \ , \quad J=-i e_i\wedge {\bar e}_{\bi} \ . 
\ee
The spin-connection $\Omega$, determined by
\be
             de_i+\Omega_i^{~k}\wedge e_k=0 \ ,
\ee
can be shown to read
\be\label{eq:def_spin_connect}
             \Omega=g^{-1}\partial g+g^{\dagger}\bar{\partial}g^{\dagger-1} \ ,
\ee
\emph{for a K\"ahler space}. This means that $\Omega$ is the connection of a holomorphic vector bundle. It is useful to split the spin-connection into its holomorphic and anti-holomorphic parts
\be
             \Omega=\mcA+\bar{\mcA} \ ,
\ee
with $\mcA=g^{-1}\partial g$ and $\bar{\mcA}=-\mcA^{\dagger}$. The holomorphy of the tangent bundle is then equivalent to the fact that
\be
             \mcF_{2,0}(\mcA)=\partial\mcA+\mcA\wedge\mcA=0 \ .
\ee
Then, the only non-vanishing piece of the tangent bundle's curvature 2-form is the $(1,1)$-component that reads
\be
             \mcR(\Omega)=\mcF_{1,1}(\mcA,\bar{\mcA})=\bar{\partial}\mcA+\partial{\bar\mcA}+\mcA\wedge\bar{\mcA}+\bar{\mcA}\wedge\mcA \ .
\ee
For future use we also note that $\mcR_i^{~j}\wedge e_j=0$.

The curvature 2-forms of Einstein-K\"ahler spaces have the particularity of satisfying the so-called hermitian Yang-Mills equation (HYM),
\be
             \mcK^{i\bj}\mcR_{i\bj}^{ab}=\frac{R}{2d_{\mathbb C}}~\delta^{ab} \ ,
\ee
with $R=-2\mcK^{i\bj}\partial_i\partial_{\bj}\ln\det (\mcK_{a\bar{b}})$ being the (constant) scalar curvature. This can also be neatly rephrased as 
\be
             J^{d_{\mathbb C}-1}\wedge\mcR=i{\mathbb 1}\frac{R}{2d^2_{\mathbb C}}J^{d_{\mathbb C}} \ ,
\ee
which, in the particular case of a Calabi-Yau $d_{\mathbb C}$-fold reads
\be\label{eq:HYM_Ricci}
             J^{d_{\mathbb C}-1}\wedge\mcR=0 \ .
\ee

A rather well studied class of non-compact CY spaces is that of CY cones, with an infinitely large number of explicit metrics being known by now  \cite{Gauntlett:2004yd,Gauntlett:2004hh,Cvetic:2005ft,Cvetic:2005vk,Lu:2005sn}. By definition, any $2d_{\mathbb C}=2(n+1)$-dimensional K\"ahler cone is a cone over a $(2n+1)$-dimensional Sasaki space, which in turn is a line bundle over a $2n$-dimensional K\"ahler base. In the following, we introduce local complex coordinates $w^i$ ($i=1,\dots,n$) for the latter
and denote its K\"ahler potential by $\mcK(w,\bw)$. The K\"ahler potential for the cone can then be written as 
\be
              \mcK_C=|z|^2e^{2\mcK(w,\bw)}\equiv\half\rho^2 \ ,
\ee
where $z\in{\mathbb C}$ vanishes at the apex of the cone. With $z=\rho e^{-\mcK}e^{i\phi}/\sqrt{2}$, we find that the metric of a K\"ahler cone,
\be
              ds^2_C=d\rho^2+\rho^2ds^2(Y) \ , \quad \rho>0 \ ,
\ee
is determined by the metric of a Sasaki space $Y$
\be
              ds^2(Y)=\eta\otimes\eta+2\mcK_{i\bj}dw^id\bw^{\bj} \ ,
\ee
where the 1-form
\be
              \eta=d\phi-i(\partial-\bar{\partial})\mcK \ ,
\ee
is the dual of a the \emph{Reeb} Killing vector field. The real coordinate $\rho$ is a radial coordinate measuring the distance to the apex of the cone. 

The 1-form $\eta$ determines the K\"ahler 2-form of the $2n$ dimensional base,
\be
              J_{EK}=-\half d\eta \ ,
\ee
as well as the K\"ahler form of the cone,
\be
              J_C=-\half d(\rho^2\eta) \ .
\ee
Likewise, the curvature of the cone is fully determined by the K\"ahler base geometry as
\be\label{eq:Ricci_cone}
              \mcR ic_C=2(n+1)J_{EK}+\mcR ic_{EK} \ ,
\ee
and
\be
              R_C=\frac{R_{EK}-4n(n+1)}{\rho^2} \ .
\ee
Therefore, imposing the Calabi-Yau condition on the cone implies that the base space is Einstein-K\"ahler with the calar curvature set by the dimensionality of the cone, 
\be
              R_{EK}=4n(n+1) \ .
\ee
Equivalently, one can state that the K\"ahler base has $U(n)$ holonomy and its curvature is HYM with 
\be\label{eq:HYM_for_basetangent}
             J^{n-1}_{EK}\wedge\mcR_{EK}=i{\mathbb 1}\frac{2(n+1)}{n}J^n_{EK} \ ,
\ee
(or $\mcK^{i\bj}\mcR_{i\bj}^{ab}=2(n+1)\delta^{ab}$).

It is also useful to have an expression for the CY cone's spin-connection $\Omega_C$ in terms of the spin-connection on the Einstein-K\"ahler base. Recall that $\Omega_C$ is determined by a rank $(n+1)$ square matrix $g$ (cf. Eq.\eqref{eq:def_spin_connect}), which in the case of a cone can be written as
\be
             g=e^{\mcK}\left(\begin{array}{cc}
              -1 &\quad 0 \\
              -2\bz\mcK_{\bar{a}} & \sqrt{2}\bz h_{\bar{a}b}
              \end{array} \right)
\ee
where $h$ is a rank $n$ matrix determining the spin-connection of the EK base. (In particular we have $(hh^{\dagger})^{\bar{b} a}=\mcK_{a\bar{b}}$.) The holomorphic part of $\Omega_C$ is then straightforward to compute
\be\label{eq:spin_connec_cone}
             \mcA_C=g^{-1}\partial g=\left(\begin{array}{cc}
              \partial\mcK &\quad 0 \\
              -\sqrt{2}e_b & \mcC_b^{~a}+ \delta_b^{~a}\partial\mcK
               \end{array} \right)
\ee
where, for convenience, we introduced the orthonormal frame $e_i=h^{\dagger}_{ia}dw^a$, and $\mcC=h^{-1}\partial h$ is the (holomorphic part of the) U($n$) spin-connection on the EK space. We should note that $\mcA_C$ is determined only up to gauge transformations which, by definition, are unitary transformations acting on $g$ from the left, leaving $gg^{\dagger}$ invariant. Since $\mcA_C$ is an SU($n+1$) connection, it must be traceless. In turn, for $\mcC$ this implies that
\be   
              \tr(\mcC)=-(n+1)\partial\mcK \ ,
\ee
and thus
\be
              \mcR ic_{EK}\equiv i\tr(\mcR_{EK})=i2(n+1)\partial\bar{\partial}\mcK=-2(n+1)J_{EK} \ .
\ee
Then, upon the use of Eq.\eqref{eq:Ricci_cone} this implies the CY condition, $\mcR ic_C=0$, to be satisfied on the CY cone as it should.

In the gauge we are using here, the curvature of the tangent bundle is straightforward to compute and gives
\be\label{eq:Tang_bundle_curvature}
              \mcR_C=\left(\begin{array}{cc}
              0 &\quad 0 \\
              0 & \mcR^{b\bar{a}}_{EK}-2(\delta^{i\bj}\delta^{b\bar{a}}+\delta^{\bar{a}\bj}\delta^{bi})e_i\wedge\bar{e}_{\bj}
              \end{array} \right)
\ee
It then follows that $J_C^n\wedge\mcR_C=0$, as expected from Eq.\eqref{eq:HYM_Ricci}.

\end{document}